\documentclass{ar-1col-S2O}
\usepackage[numbers,sort&compress]{natbib}
\usepackage{url}
\usepackage{times}
\usepackage{amsmath,amssymb}
\usepackage{comment}

\usepackage{xcolor}
\definecolor{darkblue}{rgb}{0.,0.,0.4}
\definecolor{darkred}{rgb}{0.5,0.,0.}
\definecolor{BlueViolet}{RGB}{138,43,226}
\definecolor{SkyBlue}{RGB}{30,144,255}
\definecolor{DarkGreen}{RGB}{0,100,0}

\usepackage{hyperref}
\hypersetup{
	colorlinks=true,
	linkcolor=darkblue,
	citecolor=darkblue,
	urlcolor=darkblue,
	pdftitle={A Fuzzy Sphere Journey in Critical Phenomena},
	pdfauthor={Yin-Chen He and W. Zhu}
}
\makeatletter
\AtBeginDocument{\def\@linkcolor{darkblue}\def\@citecolor{darkblue}\def\@urlcolor{darkblue}}
\makeatother

\makeatletter
\let\AR@orig@section\section
\def\section{\@ifstar{\AR@orig@section*}{\AR@caps@section}}
\newcommand{\AR@caps@section}[1]{\AR@orig@section[#1]{\MakeUppercase{#1}}}
\makeatother

\jname{Annu. Rev. Condens. Matter Phys.}
\jvol{17}
\jyear{2026}
\doi{10.1146/annurev-conmatphys-031424-020256}

\begin{document}

\markboth{He and Zhu}{A Fuzzy Sphere Journey in Critical Phenomena}

\title{A Fuzzy Sphere Journey in Critical Phenomena}

\author{Yin-Chen He$^{1,\dagger}$\thanks{$^\dagger$Current affiliation: Perimeter Institute for Theoretical Physics, Waterloo, Ontario, N2L 2Y5, Canada.} and W. Zhu$^2$
	\affil{$^1$C.N. Yang Institute for Theoretical Physics, Stony Brook University, Stony Brook, New York, USA; email: heyinchenphy@gmail.com}
	\affil{$^2$School of Science, Westlake University, Hangzhou, People's Republic of China; email: zhuwei@westlake.edu.cn}}

\begin{abstract}
This review discusses the recently proposed fuzzy sphere regularization for studying $2+1$D critical phenomena, particularly three-dimensional (3D) conformal field theory (CFT). The fuzzy sphere scheme not only offers remarkable efficiency in extracting extensive CFT data at low computational cost but also reveals unexpected connections among 3D CFT (critical phenomena), noncommutative geometry, and the quantum Hall effect. We introduce the fundamental ideas of fuzzy sphere regularization, emphasizing its role in demonstrating the state-operator correspondence of 3D CFTs on the $S^2 \times \mathbb{R}$ geometry. Additionally, we review key developments in this approach across various directions and outline potential future applications.

\end{abstract}


\maketitle

\tableofcontents

\section{Introduction}

Critical phenomena near phase transitions constitute a cornerstone of modern physics with profound implications across condensed matter physics, statistical mechanics, quantum field theory (QFT) and quantum gravity \cite{Cardy_book,Sachdev_book,Henkel_book,Vicari2002,Ma_book}. Historically, the experimental exploration of critical phenomena can be traced back three centuries to the investigation of the liquid--gas transition by Charles Cagniard de la Tour in France.
A major breakthrough came with the introduction of the Ising model \cite{Ising1925}, and its exact solution in two dimensions by Onsager \cite{Onsager1944}, conclusively demonstrating the existence of continuous phase transitions in nature. Continuous transitions and related critical phenomena exhibit a remarkable property of universality,
generally characterized by a set of critical exponents that describe nonanalytic behavior of physical observables near critical points.
For example, the liquid--gas transition shares the same universality class as the three-dimensional (3D) Ising magnetic transition.
Universality offers extensive applications of the Ising model and general theories of critical phenomena \cite{Ising_applications,RevModPhys_socialdynamics}, reaching fields such as pure mathematics, biology, earth science, deep learning, and economics, just to name a few.

Regarding the field theory description of critical phenomena, distinct universality classes correspond to distinct renormalization group (RG) fixed points \cite{Cardy_book,Ma_book,Wilson1974}.
Remarkably, at these RG fixed points, the usual Poincaré symmetry (comprising Lorentz and translation symmetries) and scale invariance often enhance into a larger spacetime symmetry—conformal symmetry—first explicitly uncovered by Polyakov through the exact solution of the two-dimensional (2D) Ising transition \cite{polyakov1970conformal}. Conformal symmetry preserves angles but not necessarily distances, signifying invariance under nonuniform RG transformations. Beyond its elegance, conformal symmetry has proven instrumental in solving critical phenomena. In two dimensions, conformal symmetry becomes infinite-dimensional \cite{Belavin1984}, rendering numerous theories exactly solvable and, thus, giving birth to conformal field theory (CFT) \cite{yellowbook}, which significantly impacts diverse areas from condensed matter physics to quantum gravity. In higher dimensions (e.g., three dimensions), conformal symmetry is less restrictive, yet modern developments such as the conformal bootstrap method \cite{Rattazzi_2008,RMP_CB,SuRychkov} have demonstrated that it remains highly effective for numerically determining precise critical exponents.

Despite widespread recognition of conformal symmetry's significance, its explicit application in studies of critical phenomena at three and higher dimensions, particularly in lattice models like the Ising model, remains limited~\cite{Weigel2000,Deng2002Conformal,Billo2013,Cosme2015Sphere}. This is primarily because lattice models typically employ torus geometries, in which the implications of conformal symmetry are unclear, thus lattice models must rely solely on scaling symmetry (finite-size scaling) to extract critical exponents. Conformal symmetry, however, becomes explicit on geometries like the sphere $S^d$ or cylinder $S^{d-1}\times \mathbb{R}$. On the cylinder, CFTs exhibit the state-operator correspondence (radial quantization)~\cite{Cardy1984,Cardy1985}, implying that quantum Hamiltonians defined on $S^{d-1}$ have eigenstates directly corresponding to scaling operators of the IR CFT. The energy gaps of these eigenstates are proportional to the scaling dimensions of the respective operators, providing a powerful framework to investigate scaling dimensions, operator product expansion (OPE) coefficients, and operator algebras. In two dimensions, studying quantum lattice models on a one-dimensional periodic chain ($S^1$) naturally leverages this geometry~\cite{Blote1986Conformal,affleck1988universal,Milsted2017,Zou2018}, but in higher dimensions, curvature poses significant challenges for conventional lattice approaches~\cite{Brower2013Lattice,Brower2021Radial}.

Parallel to developments in critical phenomena, condensed matter physics has witnessed breakthroughs in the discovery of topological phases, particularly the quantum Hall effect. Haldane proposed studying the fractional quantum Hall effect on spherical geometries through interacting fermions occupying Landau levels \cite{Sphere_LL_Haldane}. This formulation has provided invaluable insights into quantum Hall physics and topological phases broadly. Practically, Landau level models closely resemble lattice models and similarly realize phase transitions~\cite{Ippoliti2018Half,Wang2021SO5WZW}. Inspired by these advancements, Reference~\citenum{ZHHHH2022} explored three-dimensional (2+1D) critical phenomena using spherical Landau level models. Theoretically, Landau level physics corresponds to noncommutative geometry~\cite{Susskind2001}, in which spatial coordinates do not commute—analogous to the uncertainty principle of quantum mechanics. The spherical Landau level system corresponds precisely to the fuzzy (noncommutative) sphere~\cite{Hasebe2010fuzzy}, introduced initially by Madore~\cite{madore1992fuzzy}. Consequently, this approach was termed fuzzy sphere regularization~\cite{ZHHHH2022}.

Remarkably, fuzzy sphere regularization has proven highly successful for understanding 3D CFTs. Key achievements include (a) directly demonstrating conformal symmetry in the 3D Ising transition \cite{ZHHHH2022}, the $O(N)$ Wilson--Fisher (WF) model \cite{han2023o3,Lauchli_XY}, and the $SO(5)$ deconfined phase transition \cite{zhou2023so5} via state-operator correspondence; (b) accurately computing crucial quantities such as OPE coefficients \cite{hu2023operator}, the four-point correlator \cite{Han2023Conformal}, and the RG monotonic  $F$-function \cite{hu2024F}; (c) thoroughly exploring conformal line defects \cite{hu2023defect,Zhou2024gfunction} and boundaries \cite{Zhou2025surface,fuzzyHemi} in the 3D Ising transition; and (d) discovering new 3D CFTs related to Chern--Simons--matter theories \cite{zhou2024newseries3dcfts}. These advancements were facilitated by the explicit conformal symmetry on spherical geometry, and by remarkably small finite-size effects, enabling computations that previously required millions of CPU (central processing unit) hours to now be executed efficiently on standard laptops within an hour. Although many exciting results are forthcoming, this review highlights significant progress already achieved.

\section{A Lightning Review of Conformal Field Theory}
A central theme of the fuzzy sphere approach is leveraging conformal symmetry to study critical phenomena. To better facilitate the subsequent discussion, in this section we provide a brief review of CFTs, emphasizing the importance of sphere geometry as well as the state-operator correspondence formulated by Cardy in the 1980s~\cite{Cardy1984,Cardy1985}. For more comprehensive discussions on CFTs, readers are referred to textbooks such as References~\citenum{yellowbook} and~\citenum{Rychkov2016lectures}.

\subsection{Conformal Symmetry}\label{subsec:cft}
The conformal group in the \(d\)-dimensional Euclidean space is  \(\text{SO}(d+1,1)\), consisting of four types of generators~\cite{yellowbook,Rychkov2016lectures}~\footnote{For $d=2$, the conformal symmetry is infinite dimensional, and it is often called local conformal symmetry. The \(\text{SO}(d+1,1)\) conformal symmetry is sometimes called global conformal symmetry.}:
\begin{itemize}
	\item \text{Translations}: \(P_\mu = \partial_\mu\)
	\item \text{Lorentz rotations \(\text{SO}(d)\)}: \(M_{\mu\nu} = x_\nu\partial_\mu-x_\mu\partial_\nu \)
	\item \text{Dilatations}: \(D = x^\mu\partial_\mu\)
	\item \text{Special conformal transformations}: \(K_\mu = 2x_\mu x^\nu\partial_\nu - x^2\partial_\mu\).
\end{itemize}
These generators produce symmetry transformations that preserve angles but not necessarily lengths. A CFT is invariant under the conformal group and, thus, contains local operators \(\{\phi_i\}\) transforming as irreducible representations of this group. A key objective in studying CFT is understanding the properties of these local operators. The conformal generators usually do not commute, except for \([D, M_{\mu\nu}] = 0\); therefore, it is natural to consider local operators in the eigenbasis of \(D\) and \(M_{\mu\nu}\). We normally refer to these operators as scaling operators, characterized by definite scaling dimensions \(\Delta\) and Lorentz spins \(\ell\) (angular momentum of rotation). Scaling operators can be further categorized into primary operators and descendant operators: primary operators are annihilated by the special conformal transformations \(K_\mu\), whereas descendant operators can be generated by repeatedly applying translations \(P_\mu\) to the primary operators. Due to the conformal algebra, $[D, P_\mu]=P_\mu$, $P_\mu$ always increases the scaling dimension by $1$ when acting on a scaling operator. This imposes a very rigid structure to the scaling operators of a CFT.

Conformal symmetry imposes strict constraints on the correlators of scaling operators~\cite{yellowbook,Rychkov2016lectures}.  For example, the two-point function of a scalar primary in $\mathbb R^d$ is uniquely determined to be
\begin{equation}\label{eq:2pt_definition}
	\langle \phi_i(\vec x_i) \phi_j(\vec  x_j) \rangle = \frac{\delta_{ij}}{|\vec  x_{ij}|^{2\Delta_i}}.
\end{equation}
where \(\Delta_i\) is the  scaling dimension of $\phi_i$. The scaling dimensions of several operators correspond to critical exponents, such as $\eta$ and $\nu$, which are commonly measured in phase transitions.
Similarly, the three-point correlation function of scalar primaries in $\mathbb R^d$ is fixed to be
\begin{equation} \label{eq:3pt_definition}
	\langle \phi_i(\vec x_i)\phi_j(\vec x_j)\phi_k(\vec x_k)\rangle = \frac{f_{ijk}}{|\vec x_{ij}|^{\Delta_i+\Delta_j-\Delta_k}|\vec x_{jk}|^{-\Delta_i+\Delta_j+\Delta_k} |\vec x_{ik}|^{\Delta_i-\Delta_j+\Delta_k}},
\end{equation}
where \(f_{ijk}\) is the OPE coefficient, which is a theory- and operator-dependent quantity.
\( (\Delta_i,f_{ijk} )\) are called conformal data; they are the most important information for a given CFT, because the complete conformal data are able to reproduce many universal properties of the phase transition and determine the stability of fluctuations
close to a fixed point.
It is highly nontrivial to extract the conformal data of an interacting CFT. One significance of the fuzzy sphere framework is that it provides a simple way to extract the conformal data based on the state-operator correspondence.

\subsection{Conformal Field Theory on Different Geometries}

\begin{figure}[t]
    \includegraphics[width=\linewidth]{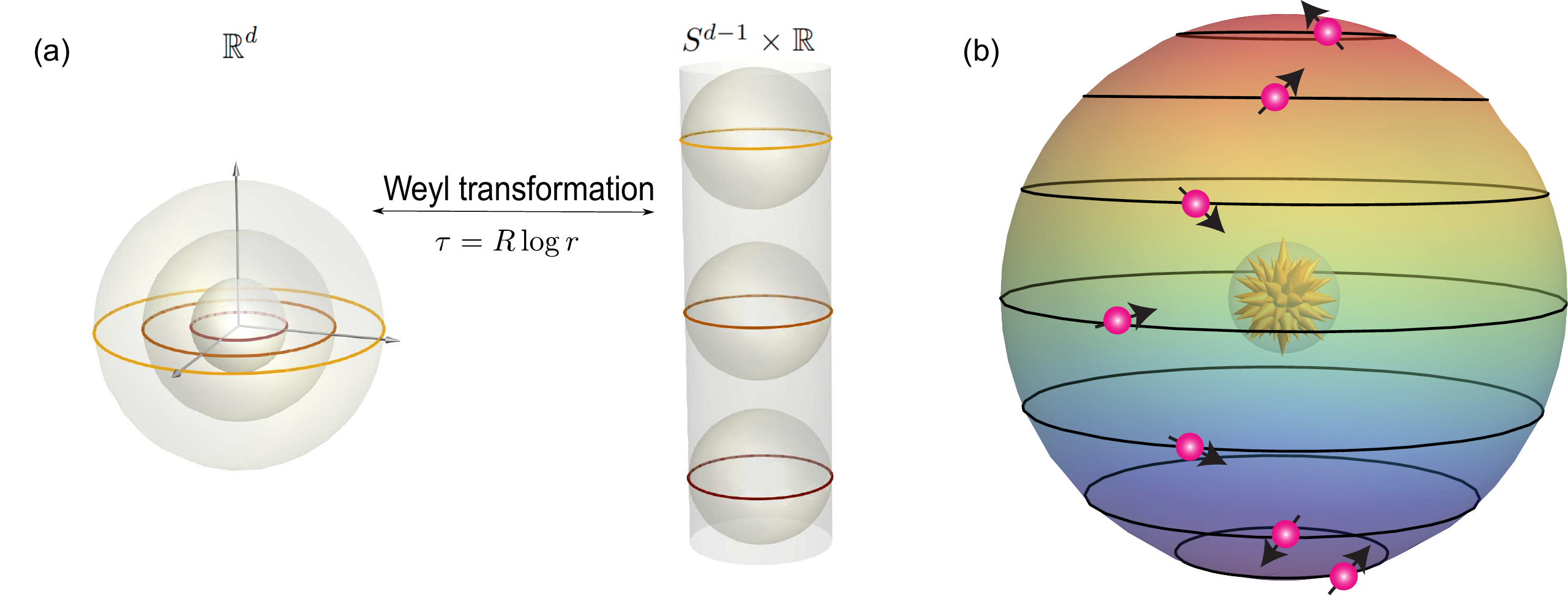}
\caption{(a) A Weyl transformation maps the Euclidean flat spacetime  $\mathbb R^d$ to the cylinder $S^{d-1}\times \mathbb R$. CFT on the cylinder has \textit{state-operator correspondence}; namely, there is a one-to-one correspondence between the Hamiltonian's eigenstates and CFT operators. (b) An illustration of a fuzzy sphere: Fermions live on the Landau level created by a $4\pi \cdot s$ monopole. The Landau orbitals are localized at different latitudes and form a spin-$s$ representation of the sphere rotation $SO(3)$. Abbreviation: CFT, conformal field theory.}
	\label{fig:radialQuan}
\end{figure}

Generically, a QFT may exhibit distinct properties when placed on different geometries. The CFT correlators defined in Equations~\ref{eq:2pt_definition} and~\ref{eq:3pt_definition} strictly apply only to the flat space $\mathbb R^d$, and they generally undergo modifications when one considers other geometries, particularly compact spaces such as spheres or tori. For example, when examining the CFT two-point correlator (Equation~\eqref{eq:2pt_definition}) on a sphere $S^d$, ambiguity arises regarding the definition of the distance $|\vec x_{ij}|$ between two points: specifically, whether it refers to the geodesic distance $R\theta_{12}$ or the linear distance $2R\sin(\theta_{12}/2)$ measured in the embedding space $\mathbb R^{d+1}$ (where $R$ is the radius of the sphere, and $\theta_{12}$ is the angular separation between the two points). A remarkable feature of CFT is that if two geometries are conformally related, then the properties of the CFT defined on these geometries can directly map onto each other~\cite{yellowbook,Rychkov2016lectures}. Importantly, $\mathbb R^d$ is conformally equivalent to both the sphere $S^d$ (through the stereographic projection) and the cylinder $S^{d-1}\times \mathbb R$ (through a Weyl transformation in Figure~\ref{fig:radialQuan}), enabling a direct inference of CFT properties on these geometries through fundamental CFT principles.
For instance, one can explicitly show that the two-point correlator on the sphere is:
\begin{equation} \label{eq:sphere2pt} \langle \phi(\vec x_i) \phi(\vec x_j) \rangle = \frac{1}{(2R\sin(\theta_{12}/2))^{2\Delta}}, \end{equation}
where the linear distance rather than the geodesic distance must be used.

In contrast, a torus geometry ($\mathbb T^d$ or $\mathbb T^{d-1}\times \mathbb R$) is not conformally equivalent to flat space $\mathbb{R}^d$ when $d\ge 3$. Consequently, even the precise form of the two-point correlator on a torus remains unknown. By utilizing scaling symmetry alone, one can show that the correlator on a torus should take the general form:
\begin{equation} \langle \phi(\vec x_i) \phi(\vec x_j) \rangle = \frac{f(\xi)}{|\vec x_{ij}|^{2\Delta}}, \quad\quad \xi = |\vec x_{ij}|/L, \end{equation}
where $L$ is the size of the torus, and $f(\xi)$ is an unknown function that is generally dependent on both the theory and the specific operator considered~\footnote{In the limit $\xi\rightarrow 0$, one has $\lim_{\xi\rightarrow 0} f(\xi)=1$.}. So for the lattice model studies on the torus, one commonly employs finite-size scaling techniques: extracting the scaling dimension $\Delta$ by fitting correlators obtained at various system sizes $L$ while keeping the ratio $\xi=|\vec x_{ij}|/L$ fixed, for example, at $\xi=1/2$. In contrast, when working on spherical geometry, one can directly extract the scaling dimension at a single fixed system size through Equation~\eqref{eq:sphere2pt}, demonstrating the power of conformal symmetry and sphere geometry~\footnote{For lattice models on a torus, the correlator still exhibits power-law behavior in the intermediate regime $1\ll|\vec x_{ij}|\ll L$. However, accurately identifying this regime in numerical simulations can be challenging, and reliably working within it typically requires very large system sizes, making the approach computationally expensive in practice.}.

Interestingly, on different geometries, the conformal generators have distinct interpretations. On the sphere $S^d$, there is a manifest $\text{SO}(d+1)$ symmetry corresponding to rotations of the sphere. The rotation generators $J_{\mu\nu}$, with indices $\mu,\nu=1,\cdots, d+1$, can be expressed in terms of the conformal generators as follows:
\begin{equation} J_{\mu\nu} = \left\{\begin{aligned} &M_{\mu\nu}, \quad\quad\quad\quad\quad \mu,\nu=1,\cdots, d, \\ &(P_\mu+K_\mu)/2, \quad\, \mu=1,\cdots, d; \nu=d+1. \end{aligned} \right. \end{equation}
On the cylinder $S^{d-1}\times \mathbb{R}$, however, one instead has a manifest $\text{SO}(d)$ symmetry corresponding to rotations of the spatial $S^{d-1}$, along with translation symmetry along the $\mathbb{R}$ direction. The rotations of $S^{d-1}$ naturally correspond to the generators $M_{\mu\nu}$, whereas the translation along $\mathbb{R}$ corresponds to the dilatation  $D$. This interpretation of the conformal generators on the cylinder is central to the state-operator correspondence discussed below.

\subsection{State-Operator Correspondence}
\label{subsec:cardy}

CFT on the cylinder $S^{d-1} \times \mathbb{R}$ is most naturally understood in the Hamiltonian formalism. One considers a quantum Hamiltonian $H$ defined on the sphere $S^{d-1}$, where the $\mathbb{R}$ direction corresponds to time evolution. The dilation generator $D$ is related to the quantum Hamiltonian via $H = D / R$, where $R$ is the radius of the sphere. Consequently, there is a one-to-one correspondence between the eigenstates of $H$ and the scaling operators of the CFT; it is a relationship known as the \emph{state-operator correspondence} \cite{Cardy1984,Cardy1985}. More importantly, the energy of each eigenstate corresponds to the scaling dimension of the associated CFT operator, given by $E = \Delta / R$. Each eigenstate also carries a definite $SO(d)$ angular momentum $\ell$, which corresponds to the Lorentz spin of the associated CFT operator. In practical studies of a quantum Hamiltonian on the sphere, the Hamiltonian may have an arbitrary normalization and an additive constant shift. As a result, the eigenstate energy gaps are proportional to the scaling dimensions $\Delta_n$ of the CFT operators:
\begin{equation}\label{eq:energy_to_scalingdim}
	\delta E_n = E_n - E_0 = \frac{v}{R} \Delta_n.
\end{equation}
Using this formula, one can directly compute scaling dimensions without doing any of the data fitting that is commonly done in the lattice simulation of critical phenomena.

Let us further explain the state-operator correspondence using the language of Weyl transformations. This perspective is particularly useful for studying defect and boundary CFTs, which we also discuss in this review.
We begin with $\mathbb{R}^d$ and choose an origin, and then foliate $\mathbb{R}^d$ into spheres $S^{d-1}$ of varying radii, as illustrated in Figure~\ref{fig:radialQuan}. Clearly, these spheres exhibit a manifest $SO(d)$ rotational symmetry, whereas the dilatation maps spheres of different radii onto one another. In this setup, the state-operator correspondence states that operators inserted at the origin correspond to states defined on a sphere. This formulation of the correspondence is also referred to as radial quantization.
Now, we map $\mathbb{R}^d$, parametrized by $(\mathbf \Omega,r)$, onto the cylinder $S^{d-1} \times \mathbb{R}$, parametrized by $( \mathbf \Omega,\tau)$, via the Weyl transformation $\tau = R \log r$. Under this mapping, dilatations along the radial direction in $\mathbb{R}^d$ correspond to translations along the $\mathbb{R}$ direction on the cylinder. The CFT operator inserted at $\tau=-\infty$ is in one-to-one correspondence with the state defined on the sphere $S^{d-1}$.

\section{Fuzzy Sphere Regularization}
As discussed above, CFTs on the cylinder $S^{d-1} \times \mathbb{R}$ enjoy several advantages owing to their conformal equivalence to $\mathbb{R}^d$. In particular, the state-operator correspondence provides a convenient framework for extracting conformal data, such as scaling dimensions. However, a nonperturbative observation of the state-operator correspondence for 3D CFTs had long been elusive, as a regular lattice cannot be embedded into the curved $S^2$. The recently developed approach of fuzzy sphere regularization \cite{ZHHHH2022} circumvents this obstacle by formulating the theory on a fuzzy (noncommutative) sphere rather than a discretized sphere.

It is worth mentioning that QFT on the fuzzy sphere has been explored before~\cite{Grosse1995,Madore2001Scaling,FuzzyQFTnumerics}, and it has long been regarded as similar to noncommutative field theory in flat space~\cite{noncommuQFT}, which exhibits UV-IR mixing and, therefore, does not reproduce standard QFT or CFT behavior. The prevailing intuition is that moving from a commutative to a noncommutative space may lead to a loss of locality, ultimately preventing the theory from behaving as a conventional QFT. Thus, a key aspect of constructing QFTs on the fuzzy sphere—or on noncommutative geometries more generally—is ensuring that locality is maintained.  The new fuzzy sphere approach~\cite{ZHHHH2022}, however, is inspired by quantum Hall physics on the sphere, which was first studied by Haldane~\cite{Sphere_LL_Haldane}. The central idea is to study interacting fermions confined to the lowest Landau level (LLL) on the sphere. This quantum mechanical system has a well-defined notion of locality and can realize 3D CFTs as phase transitions, in a manner analogous to how lattice models realize CFTs.

In general, fuzzy sphere models are built of fermions with an extra (pseudo-)spin degree of freedom. And it is the spin degree of freedom that goes through the phase transitions, whereas the charge degree of freedom is always gapped. One natural way to insure that the charge is gapped is to choose integer filling of fermions such that the charges are forming an integer quantum Hall state. This is again analogous to lattice models such as the Hubbard model in the deep Mott regime, whose effective model is a pure spin model, i.e., a Heisenberg model~\footnote{However, the fuzzy sphere model cannot be written as a pure spin model because the charge fluctuation is gapped but not completely frozen.}.

\subsection{Framework}

The fuzzy sphere regularization considers an interacting quantum mechanical system with many particles (e.g., fermions) living on the sphere $S^2$ in the presence of a magnetic monopole at the origin. Generically, we can describe the system by a Hamiltonian,
\begin{equation}\label{eq:generalham}
H= H_{kin}+ H_{int}= \frac{1}{2M_eR^2} \int d^2\mathbf{x}\, \psi^\dag(\mathbf{x}) (\partial_\mu + i A_\mu)^2 \psi(\mathbf{x}) + H_{int}.
\end{equation}
Here, $\psi(\mathbf{x})$ is the fermion field that satisfies the standard fermion anticommuting relations, $R$ is the radius of the sphere, and $M_e$ is the mass of fermions. $A_\mu$ describes the gauge field generated by the $4\pi s$ monopole ($2s\in \mathbb{Z}$). $H_{int}$ is the interaction term, which can take the form of density--density interaction, and is discussed in detail later. This model is a local model as long as the interaction $H_{int}$ is local.

We consider the limit  $H_{kin}\gg H_{int}$ so that we can analyze the kinetic term first.  $H_{kin}$ can be solved directly, and its eigenstates are forming quantized spherical Landau levels with energies~\cite{Sphere_LL_Haldane,Sphere_LL_Greiter}:
\begin{align}\label{eq:eq:LLeigenvalues}
	E_n= \frac{1}{2M_e R^2}[n(n+1)+(2n+1)s]
\end{align}
where $n=0, 1, 2, \cdots$  is the Landau level index, and the $(n+1)_\textrm{th}$ Landau level is $(2s+2n+1)$-fold degenerate with these degenerate states forming spin-$n+s$ irreducible representation of sphere rotation $SO(3)$. Especially, for the LLL, the single-particle eigenstates (Landau orbitals) are described by the wave functions  called monopole harmonics~\cite{WuYangmonopole} (also called spin-weighted spherical Harmonics)
\begin{equation} \label{eq:orbitals}
	Y^{(s)}_{s,m} (\theta, \varphi)=\mathcal N_m e^{im\varphi} \cos^{s+m}\left(\frac{\theta}{2}\right)\sin^{s-m}\left(\frac{\theta}{2}\right),
\end{equation}
with $m=-s, -s+1, \cdots, s$ and $\mathcal N_m = \sqrt{\frac{(2s+1)!}{4\pi(s+m)!(s-m)!}}$, and $(\theta, \varphi)$ is the spherical coordinate. These wave functions form the spin-$s$ representation of $SO(3)$, and they are localizing at different latitudes of the sphere as shown in Figure~\ref{fig:radialQuan}.

In the limit $H_{kin}\gg H_{int}$, the energy gap between Landau levels is much greater than the interactions. So if the LLL is not completely filled, we can just consider the LLL $n=0$ and project out the other Landau levels. This process is usually called LLL projection in the quantum Hall community \cite{Jain_2007}.
Technically, this can be done by replacing the fermion field $\psi(\mathbf{x})$ by operators on the LLL,
\begin{equation}
 \psi(\mathbf{x}) = \frac{1}{\sqrt{2s+1}}  \sum_{m=-s}^s \bar Y^{(s)}_{s,m}(\theta, \varphi)  c_m.
\end{equation}
Here, $ c_m$ stands for the annihilation operator of Landau orbital $m$, which follows the standard fermion anticommutation relation~\footnote{After the LLL projection, the anticommutation relation of $\psi(\mathbf{x})$ gets modified.}. And we replace the radius of sphere $R$ with $\sqrt{2s+1}$.

The Hilbert space of the system is the Fock space of fermions living on the Landau orbitals. Any field can be  written as the second quantized operators acting on the Hilbert space. For example,  the density field $n(\mathbf{x}) = \psi^\dag(\mathbf{x}) \psi(\mathbf{x})$ can be written as
\begin{equation}
	n(\mathbf{x}) = \frac{1}{2s+1} \sum_{m_1, m_2} c^\dag_{m_1} c_{m_2}  Y^{(s)}_{s,m_1}(\theta, \varphi) \bar Y^{(s)}_{s,m_2}(\theta, \varphi) ,
\end{equation}
and the density--density interaction $H_{int}= \int d^2  \mathbf x_a d^2  \mathbf x_b \; U(\mathbf x_a -\mathbf x_b) n (\mathbf x_a)  n (\mathbf x_b)$ can be written as,
\begin{align}\label{eq:generic_LLL_dd}
	H_{int} &=  \int d^2\mathbf x_a d^2\mathbf x_b \, U(\mathbf x_a -\mathbf x_b) n(\mathbf x_a)  n(\mathbf x_b)  \nonumber \\
	& = \sum_{m_1, m_2, m_3, m_4} V_{m_1, m_2, m_3, m_4} \, c^\dag_{m_1} c^\dag_{m_2} c_{m_3}  c_{m_4},
\end{align}
where $V_{m_1, m_2, m_3, m_4}$ encodes the details on the interaction. $V_{m_1, m_2, m_3, m_4}$ can be further expanded using the so-called Haldane pseudopotential  $V_l$ \cite{Sphere_LL_Haldane}, corresponding to the two-fermion scattering in the spin-$2s-l$ channel.

Here, we present several remarks in order. First, the LLL on the sphere indeed corresponds to the fuzzy sphere, a fundamental example of noncommutative geometry in mathematics. To provide an intuitive understanding of the emergence of the fuzzy sphere, we consider the projection of the coordinates of a unit sphere, denoted as $\vec{x} = (\sin\theta\cos\varphi, \sin\theta\sin\varphi, \cos\theta)$. After the projection, these coordinates become three $(2s+1) \times (2s+1)$ matrices, given by
\begin{equation}
    (\vec{X})_{m_1,m_2} = \int \sin\theta\,\mathrm{d}\theta\,\mathrm{d}\varphi \, \vec{x}\, \bar{Y}^{(s)}_{s, m_1}(\theta,\varphi) Y^{(s)}_{s, m_2}(\theta,\varphi).
\end{equation}
These matrices satisfy the following commutation relations:
\begin{equation}
    [\mathbf{X}^\mu, \mathbf{X}^\nu] = \frac{1}{s+1} i\epsilon^{\mu\nu\rho}\mathbf{X}_\rho, \quad \mathbf{X}_\mu\mathbf{X}^\mu = \frac{s}{s+1} \mathbf{1}_{2s+1}.
\end{equation}
The fact that the three projected coordinates satisfy the $\mathrm{SO}(3)$ algebra formally defines a fuzzy sphere \cite{madore1992fuzzy}. Notably, in the limit $s \to \infty$, the fuzziness disappears, and the commutative unit sphere is recovered. This limit also corresponds to the continuum limit in our fuzzy sphere regularization.

Second, the final model we study is defined on the LLL, which has a finite-dimensional Hilbert space (when $s$ is finite). In the LLL basis, the Hamiltonian exhibits all-to-all interactions, making locality less transparent. However, it is important to note that the original Hamiltonian (Equation~\ref{eq:generalham}) remains perfectly local as long as the interactions are local in real space. One can view the LLL projection as a mathematical technique, fully justified once we choose the interaction strength to be much smaller than the energy gaps between Landau levels.

Third, QFTs and critical phenomena require proper regularization schemes to manage divergences while preserving physical symmetries. For example, the lattice regularization discretizes the lattice but breaks the continuous spatial symmetries to discrete ones.
In contrast, fuzzy sphere regularization balances the symmetry and regularization by replacing spacetime with a \textit{noncommutative manifold} that retains exact rotational symmetry.

In summary, the fuzzy sphere model we study consists of a fermionic Hamiltonian encompassing $2s+1$ Landau orbitals with $SO(3)$-invariant spatial interactions. The advantage of this framework is that it strikes a balance between preserving symmetry and implementing regularization while providing a direct way to realize 3D CFTs on the sphere, in which the state-operator correspondence can be examined.  Furthermore, one can design microscopic models with various global symmetries and interactions to realize different universalities corresponding to distinct CFTs. These models can be analyzed using numerical diagonalization and other computational methods, as discussed below.

\subsection{Open-Source Package: FuzzifiED}

As introduced in this review, fuzzy sphere regularization offers significant advantages in studying 3D CFTs, opening an avenue for exploring a wide range of physics with minimal computational cost.   To bridge the gap in performing numerical computations on the fuzzy sphere, Zhou has developed an open-source software package, \texttt{FuzzifiED} \cite{zhou2025fuzzifiedjuliapackage}, for reference and practical use. \texttt{FuzzifiED} is a Julia package designed to simplify numerical calculations for researchers interested in the fuzzy sphere framework. It facilitates exact diagonalization as well as density matrix renormalization group (DMRG) calculations (using \texttt{ITensor} \cite{itensor}). The package supports a broad class of fermionic and bosonic models and has provided sample codes for nearly all the papers that have published on the fuzzy sphere regularization.
 (Detailed documentation and instruction for using the package is available online at: documentation, \url{https://docs.fuzzified.world}; source code, \url{https://github.com/FuzzifiED/FuzzifiED.jl}.)

\section{Three-Dimensional Ising Conformal Field Theory on Fuzzy Sphere}
In this section, we use the 3D Ising criticality as the primary example to demonstrate the fuzzy sphere scheme in more detail~\cite{ZHHHH2022}. The next section surveys more recent progress.

\subsection{Hamiltonian and Phase Transition}
We consider fermions $\Psi^\dagger(\mathbf x) = (\psi^\dagger_{\uparrow}(\mathbf x), \psi^\dagger_{\downarrow}(\mathbf x))$ carrying an isospin degree of freedom ($\sigma=\uparrow,\downarrow$) on sphere subject to a magnetic monopole with $4\pi s$ flux. Using these fermion fields, we can construct density fields that are analogs of spin operators in lattice models,
	\begin{equation}
		n^\alpha (\mathbf x) = ( \hat\psi^\dag_{\uparrow}(\mathbf x), \, \hat\psi^\dag_{\downarrow}(\mathbf x) ) \, \sigma^\alpha \left(\begin{matrix}
			\hat\psi_{\uparrow}(\mathbf x) \\
			\hat\psi_{\downarrow}(\mathbf x)
		\end{matrix} \right),
	\end{equation}
	with $\sigma^{x,y,z}$ being Pauli matrices, $\sigma^0=I_{2\times 2}$. The Hamiltonian is written using fields (here, we omit the kinetic term),
	\begin{align} \label{eq:ham_ising}
		H &=  \frac{1}{2} \int_{S^2} d^2  \mathbf x_a d^2  \mathbf x_b U(\mathbf x_a-\mathbf x_b)\left[n^0(\mathbf x_{a})n^0(\mathbf x_{b})- n^z(\mathbf x_{a})n^z(\mathbf x_{b}) \right] - h \int_{S^2}  d^2  \mathbf x \, n^x(\mathbf x),
	\end{align}
	Here, the term $n^z(\mathbf x_{a})n^z(\mathbf x_{b})$ is the spherical version of Ising ferromagnetic interaction, whereas the last term is the transverse field term.
    $U(\mathbf x_a-\mathbf x_b)$ describes the form of density--density interaction, and we choose it to be a local form $U(\mathbf x_a-\mathbf x_b) = g_0\delta(\mathbf x_a-\mathbf x_b) + g_1\nabla^2\delta(\mathbf x_a-\mathbf x_b)$. Two parameters, $g_0$ and $g_1$, offer additional degrees of freedom to fine-tune the contribution of irrelevant operators and to minimize the finite-size effects.

	Next, we project the Hamiltonian into the LLL (fuzzy sphere). We obtain
	\begin{align}\label{eq:LLL_dd}
		H &= \sum_{m_1, m_2, m_3, m_4} V_{m_1, m_2, m_3, m_4} \,  c^\dag_{m_1,\uparrow} c^\dag_{m_2,\downarrow} c_{m_3,\downarrow} c_{m_4,\uparrow} -h \sum_{m=-s}^s \mathbf{c}_{m}^\dag \sigma_x \mathbf{c}_{m},
	\end{align}
	where  $\mathbf c^\dag_{m} = (c^\dag_{m,\uparrow}, c^\dag_{m,\downarrow})$ is the spinful fermion creation operator on the $m_\textrm{th}$ Landau orbital.
	Following the strategy of quantum Hall community, we connect the interaction parameter $V_{m_1,m_2,m_3,m_4}$ to the Haldane pseudopotentials and use pseudopotential parameters $V_0$ and $V_1$ to tune the form of interaction \cite{Sphere_LL_Haldane}.
	We consider the half-filling case with the LLL filled by $N=2s+1$ electrons.

	When $h=0$ (interaction dominated case), the ground state is an Ising ferromagnet that spontaneously breaks $\mathbb{Z}_2$ symmetry. In the quantum Hall literature, this phase is called quantum Hall ferromagnetism \cite{Sondhi1993,Girvin2000}. The two-fold degenerate ground states are $|\Psi_{\uparrow}\rangle = \prod_{m=-s}^s c^\dag_{m\uparrow} |0\rangle $ and $|\Psi_{\downarrow}\rangle = \prod_{m=-s}^s c^\dag_{m\downarrow} |0\rangle $. When $h \rightarrow \infty$, the ground state is a trivial paramagnet that preserves Ising symmetry, $|\Psi_{x}\rangle = \prod_{m=-s}^s (c^\dag_{m\uparrow}+c^\dag_{m\downarrow}) |0\rangle $. Therefore, by tuning $h$, we expect a $2+1D$ Ising transition at some critical field $h_c$.
	In the entire phase diagram, electric charges of fermions are gapped, whereas the Ising spins of fermions are the degrees of freedom that go through the phase transitions.
	All the gapless modes (in the low-energy regime) at the phase transition are from charge-neutral spin degrees of freedom.

Our Hamiltonian has a number of symmetries:
	\begin{enumerate}
		\item Ising $\mathbb{Z}_2$ symmetry: $(c_{m\uparrow}, c_{m\downarrow}) \rightarrow (c_{m\downarrow}, c_{m\uparrow}) = (c_{m\uparrow}, c_{m\downarrow})\sigma^x $.
		\item $SO(3)$ sphere rotation symmetry: $c_{m,\sigma}$, $m=-s,\cdots, s$, form the spin-$s$ representation of $SO(3)$.
		\item Particle--hole symmetry:  $(c_{m,\uparrow}, c_{m,\downarrow}) \rightarrow (c_{m,\downarrow}^\dag, -c_{m,\uparrow}^\dag)$. It acts as an improper $\mathbb{Z}_2$ of $O(3)$, and it can be identified as the spacetime parity of the 3D Ising CFT.
	\end{enumerate}
These symmetries are helpful for numerical simulations and also useful for identifying the connections between our fuzzy sphere model and the IR CFT.

To study various conformal properties of the 3D Ising transition, we must first determine the phase transition point. Several methods can be used to locate the critical point, particularly by leveraging the advantages of the cylinder geometry. The original paper~\cite{ZHHHH2022} employed a traditional approach—finite-size scaling of the order parameter. The Ising order parameter is defined as
\begin{equation}
    M = \sum_{m=-s}^s \mathbf{c}^\dag_m \sigma^z \mathbf{c}_m.
\end{equation}
At the critical point, we expect $\langle M^2 \rangle \sim R^{4-2\Delta_\sigma} = N^{2- \Delta_\sigma}$, where $\Delta_\sigma \approx 0.5181489$ \cite{RMP_CB,Ising_CB,StressTensorBoot} is the scaling dimension of the primary operator $\sigma$. A crossing-point analysis based on numerical simulations for various system sizes yields the transition point $h_c$.

Beyond this traditional method, it is preferable to use approaches that directly exploit conformal symmetry on the cylinder. For instance, the equal-time two-point correlator should take the conformal form given in Equation~\eqref{eq:sphere2pt}. Thus, we can identify the phase transition point by finding interaction parameters at which the correlator best matches the expected conformal behavior~\cite{Han2023Conformal,Hofmann2024}, as we discuss below in Section~\ref{sec:operators}. Another approach relies on the state-operator correspondence: We can either search for interaction parameters that exhibit the highest-quality conformal symmetry in their operator spectrum~\cite{zhou2024newseries3dcfts}, or employ conformal perturbation theory, as shown in Reference~\citenum{Lauchli_CPT}. Notably, these two methods eliminate the need for finite-size scaling, as a single sufficiently large system can already yield accurate results~\footnote{In practice, finite-size extrapolation can still be performed to further refine the results, in particular when the system size is not huge.}.

In practice, when simulating small system sizes, finite-size effects are unavoidable. A common source of these effects is the coupling of irrelevant operators, which are typically present in microscopic models. Adjusting parameters in the model generally alters the coupling strength of irrelevant operators and, consequently, the magnitude of finite-size effects.  In the following section, numerical data are presented for a specific interaction parameter set $(V_0 = 4.75, V_1 = 1.0, h_c = 3.16)$, where finite-size effects are found to be minimal~\cite{ZHHHH2022}.

\subsection{Conformal Tower and Operator Spectra} \label{sec:stateop}

We now discuss the state-operator correspondence on $S^2 \times \mathbb{R}$, in which the eigenstates of the quantum Hamiltonian are in one-to-one correspondence with the scaling operators of the CFT. In particular, the energy gaps of the quantum Hamiltonian are proportional to the scaling dimensions of CFT operators (see Equation~\eqref{eq:energy_to_scalingdim}). This allows us to directly extract conformal data by computing the low-lying eigenstates of our fuzzy sphere Hamiltonian at the critical point~\cite{ZHHHH2022}.

To match the energy spectrum of the Ising transition with the operator spectrum of the 3D Ising CFT, we must first rescale the energy levels to eliminate the nonuniversal numerical factor $v/R$ in Equation~\eqref{eq:energy_to_scalingdim}. Several methods based on conformal symmetry can be used for this purpose. One natural choice is to utilize the energy-momentum tensor $T_{\mu\nu}$, a conserved operator present in any local CFT. In a 3D CFT, $T_{\mu\nu}$ is a Lorentz spin $\ell = 2$ operator that is a singlet under global symmetries, with a fixed scaling dimension $\Delta_T = 3$. Our model respects exact $SO(3)$ Lorentz rotation, Ising $\mathbb{Z}_2$, and spacetime parity symmetries, ensuring that each eigenstate carries well-defined quantum numbers $(\mathbb{Z}_2, P, \ell)$ corresponding to these symmetries. The energy--momentum tensor is identified as the lowest-energy state in the $(\mathbb{Z}_2 = 1, P = 1, \ell = 2)$ sector. We rescale the full spectrum by setting the energy of the energy--momentum tensor to exactly $\Delta_T = 3$. This automatically aligns the entire (low-lying) spectrum with the scaling dimensions of the CFT’s scaling operators—both primaries and descendants—up to finite-size corrections.

\begin{figure*}
	\centering
	\includegraphics[width=0.95\textwidth]{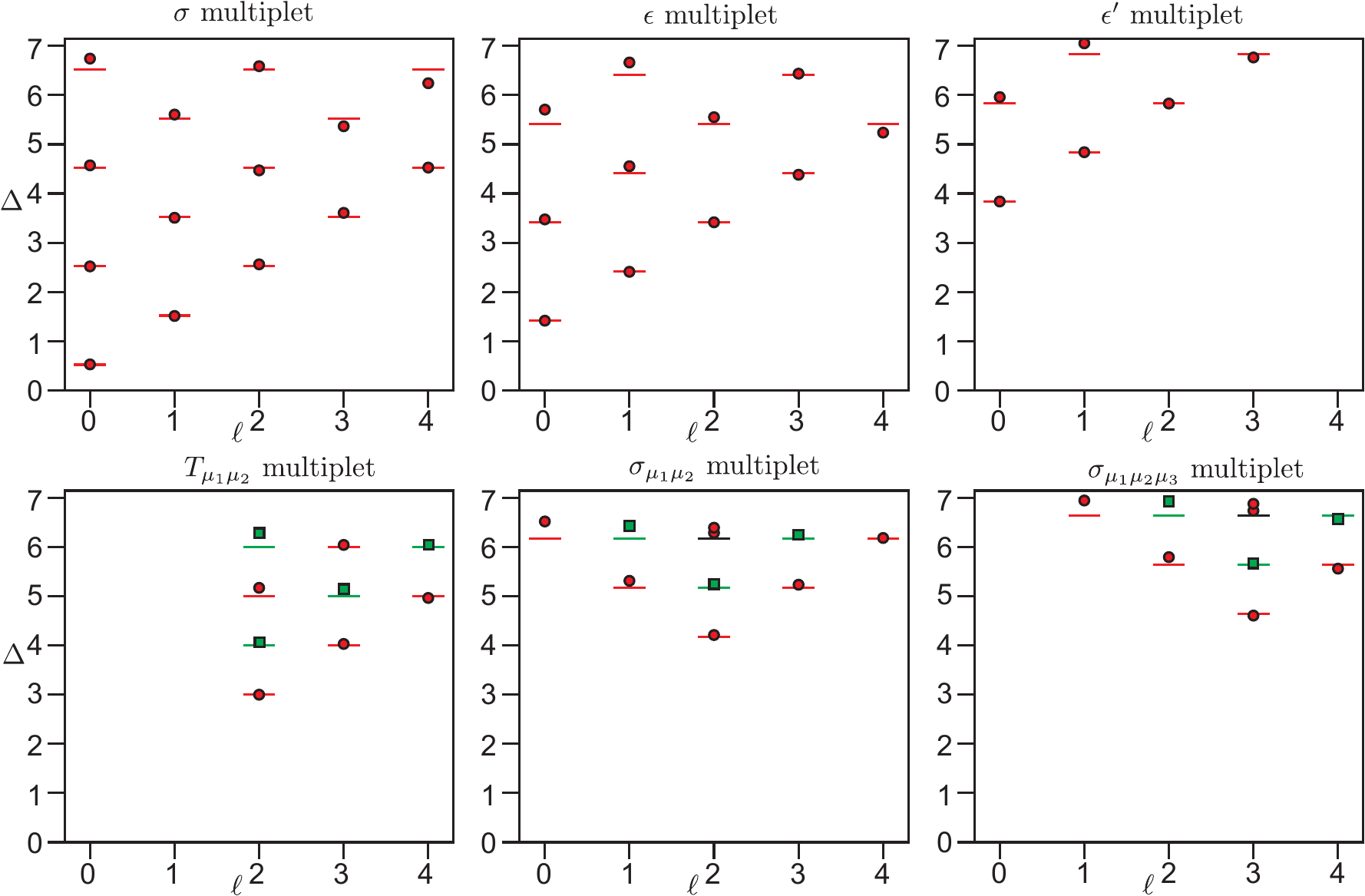}
\caption{Conformal multiplet of several low-lying primary operators: scaling dimension $\Delta$ versus Lorentz spin $\ell$. We plot conformal bootstrap data with lines: lines in red are parity even, nondegenerate operators;  lines in green are parity odd, nondegenerate operators; and lines in black are parity even, two-fold degenerate operators. Symbols are our numerical data of parity even (red circle) and odd (green square) operators.
			The discrepancy is typically more significant for the larger $\Delta$.
    Figure adapted from Reference~\citenum{ZHHHH2022} (CC BY 4.0).
		}
	\label{fig:multiplet}
\end{figure*}

The spectrum can then be organized into conformal multiplets, i.e., irreducible representations of the conformal group. Each conformal multiplet consists of a primary state along with its descendants, following a rigid structure dictated by the conformal algebra. Specifically, the primary state is annihilated by the special conformal generator $K_\mu$, whereas the descendants are generated by acting on the primary state with the translation generator $P_\mu$ one or more times.  More importantly, each multiplet exhibits a tower-like structure, often referred to as a conformal tower, characterized by integer spacings between the scaling dimensions of the primary state and its descendants. Figure~\ref{fig:multiplet} illustrates the conformal multiplets of several representative primary operators, clearly revealing the expected conformal tower structure. This provides a direct and unambiguous demonstration of the emergent conformal symmetry of the 3D Ising transition, thus verifying the conjecture proposed by Polyakov half a century ago \cite{polyakov1970conformal}.

Beyond matching the scaling dimensions of the conformal tower, one can take a further step by explicitly constructing the conformal generators in the fuzzy sphere model and examining the conformal algebra. This has been thoroughly investigated in Reference~\citenum{fardelli2024}, which we review in Section~\ref{sec:generator}.

Table~\ref{tab:primary} lists all the primary operators Reference~\citenum{ZHHHH2022} has identified on a given system with $N=16$ fermions.
Reference~\citenum{ZHHHH2022} has found 12 parity-even primary operators besides the energy--momentum tensor, and all of them  have less than $1.6\%$ discrepancy from the state-of-the-art conformal bootstrap calculation \cite{Ising_CB,RMP_CB,StressTensorBoot}.
Even more remarkably, all $70$ low-lying states with $\ell\le 4$ perfectly  match the $3D$ Ising CFT's spectrum (containing both primary and descendant operators), without any extra or missing operators below scaling dimension $\Delta=7$.
For example, low-lying spectrum in the $\mathbb{Z}_2-$odd sector are $\Delta=0.524289, 1.509417, 2.517221, \cdots$, and they respectively correspond to the primary operator $\sigma$ and its descendant operators $\partial_\mu \sigma$, $\square \sigma$, $\cdots$. (Small deviations from the integer spacing are due to the finite-size effect.)
The numerical accuracy is unexpectedly high, particularly given that it is from a small system size ($N=16$ total fermions): Around 10 operators have relative numerical error around $3\%\sim 5.5\%$, and the rest of them have relative numerical error smaller than $3\%$~\footnote{The results can be further improved by using conformal perturbation theory~\cite{Lao2023}.}.
Additionally, the symmetry features of these operator spectra are nicely consistent with the predictions of 3D Ising CFT as detailed in Reference~\citenum{ZHHHH2022}.
Furthermore, the substantial quantity of operator content obtained by the fuzzy sphere surpasses that from traditional methods. For example, in all previous lattice model studies, only several primary operators ($\sigma$, $\epsilon$, and $\epsilon'$) were found, and their scaling dimensions are related to the critical exponents $\eta$, $\nu$, and $\omega$~\cite{Hasenbusch2010,Landau2018}. Another example is that the fuzzy sphere identifies two previously unknown (parity-odd) primary operators $\sigma^{P-}$ and $\epsilon^{P-}$.
Furthermore, it is also worth noting that the fuzzy sphere obtains these conformal data at a very low cost.
Even for a much smaller system size, such as $N = 4$ fermions, one can already obtain six primary operators and their descendants with remarkably accurate scaling dimensions.

\begin{table*}
	\setlength{\tabcolsep}{0.2cm}
	\renewcommand{\arraystretch}{1.4}
	\centering
		\caption{Low-lying primary operators of three-dimensional Ising conformal field theory identified via state-operator correspondence on a fuzzy sphere with $N=16$ fermions$^a$.} \label{tab:primary}
	\begin{tabular}{cccccccc|c} \hline\hline
		& $\sigma$ & $\sigma'$  & $\sigma_{\mu_1 \mu_2}$ & $\sigma'_{\mu_1 \mu_2}$ & $\sigma_{\mu_1 \mu_2 \mu_3}$ & $\sigma_{\mu_1 \mu_2 \mu_3 \mu_4}$ & & $\sigma^{P-}$\\
		Bootstrap  & 0.518 & 5.291 & 4.180  & 6.987  & 4.638 & 6.113 && NA \\
		Fuzzy sphere & 0.524 & 5.303 & 4.214 & 7.048 & 4.609  & 6.069 && 11.191 \\
		\hline
		& $\epsilon$ & $\epsilon'$  & $\epsilon''$  & $T_{\mu\nu}$ & $T'_{\mu\nu}$ & $\epsilon_{\mu_1\mu_2\mu_3\mu_4}$ & $\epsilon'_{\mu_1\mu_2\mu_3\mu_4}$ & $\epsilon^{P-}$ \\
		Bootstrap  & 1.413 & 3.830  & 6.896 & 3 & 5.509 & 5.023 &  6.421 & $\le 10.93$  \\
		Fuzzy sphere & 1.414 & 3.838 & 6.908  & 3 & 5.583 & 5.103 &  6.347 & 10.014 \\
		\hline\hline

		\end{tabular}
		\par\smallskip
		\begin{minipage}{0.98\textwidth}
			\footnotesize
			Abbreviation: NA, not available.

			$^a$The operators in the first and second row are $\mathbb{Z}_2$ odd and even operators, respectively. We highlight that two new parity-odd primary operators $\sigma^{P-}$ and $\epsilon^{P-}$ are found. The conformal bootstrap data are from References~\citenum{RMP_CB} and~\citenum{StressTensorBoot}. Table adapted with permission from Reference~\citenum{ZHHHH2022}; copyright 2023 American Physical Society.
		\end{minipage}
\end{table*}

\subsection{Operators, Correlators, and Operator Product Expansion Coefficients}\label{sec:operators}

Beyond the energy spectrum, it is also important to study operators in the fuzzy sphere model. As we review in this section, operators not only provide access to key information such as the OPE coefficients but also help establish the connection with standard QFT.

\begin{figure}[t]
	 \includegraphics[width=\linewidth]{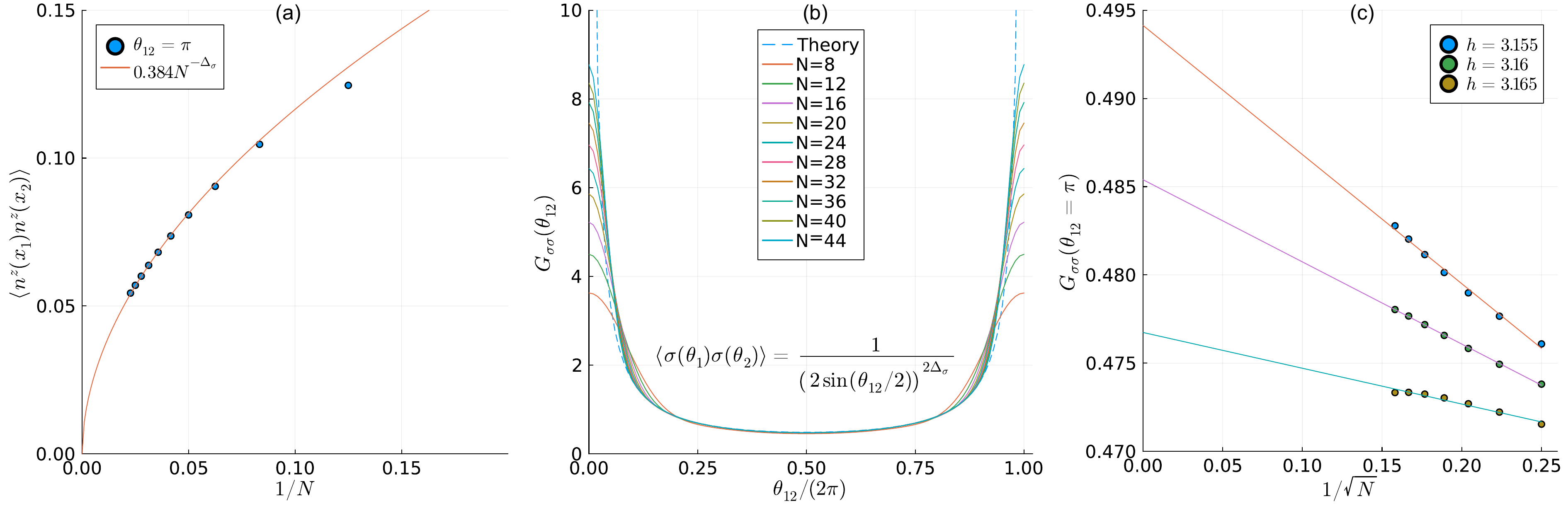}
\caption{Two-point correlator of $n^z\sim \sigma$. (a) The correlator at the antipodal points ($\theta_{12}=\pi$) shows the expected $N^{-\Delta_{\sigma}}$ scaling. (b) The dimensionless two-point correlator $G_{\sigma\sigma}(\theta_{12})$ shows excellent agreement with the theoretical expectation. (c) $G_{\sigma\sigma}(\theta_{12}=\pi)$ at and off the critical point. Figure adapted from Reference~\citenum{Han2023Conformal}.
\label{fig:correlator}
}

     \end{figure}

In the fuzzy sphere model, computations are performed in the orbital space of the LLL, but one can also consider operators (quantum fields) and other physical quantities defined in real space. The most fundamental fields are the density fields $n^\alpha(\mathbf{x})$, given by \cite{hu2023operator}
\begin{equation}
    n^\alpha (\mathbf{x})  = \frac{1}{2s+1} \sum_{m_1,m_2=-s}^s ( c^\dag_{m_1,\uparrow}, \, c^\dag_{m_1,\downarrow} ) \, \sigma^\alpha \left(\begin{matrix}
        c_{m_2,\uparrow} \\
        c_{m_2,\downarrow}
    \end{matrix} \right) Y^{(s)}_{s,m_1}(\theta, \varphi) \bar Y^{(s)}_{s,m_2}(\theta, \varphi),
\end{equation}
which closely resemble quantum fields in standard QFT. From a symmetry perspective, the density fields $n^{x,y,z}(\mathbf{x})$ correspond to familiar fields in the 3D Ising CFT~\footnote{$n^x(\mathbf{x})$ also contains a contribution from the identity operator, which must be subtracted.},
\begin{equation}
    n^x(\mathbf{x}) \sim \epsilon(\mathbf{x}), \quad\quad n^y(\mathbf{x}) \sim \partial_t \sigma(\mathbf{x}), \quad\quad \text{and} \quad n^z(\mathbf{x}) \sim \sigma(\mathbf{x}).
\end{equation}
These identifications of fields imply that, in the limit $s \to \infty$, computing the correlators of a density field yields the corresponding CFT correlator. For instance, the equal-time two-point correlator behaves as follows (here and throughout, we consider operators at $\tau = 0$ unless otherwise specified):
\begin{equation}
    \langle n^z(\mathbf{x}_1) n^z(\mathbf{x}_2) \rangle \sim \frac{1}{\left(2\sqrt{2s+1}\sin(\theta_{12}/2)\right)^{2\Delta_\sigma}},
\end{equation}
where $\sqrt{2s+1}$ is equivalent to the sphere radius $R$, and $\theta_{12}$ is the relative angle between the two points. Figure~\ref{fig:correlator}(a) confirms that the antipodal ($\theta_{12}=\pi$) correlator indeed follows the expected $N=2s+1$ dependence.

Furthermore, we can define a dimensionless two-point correlator using the state-operator correspondence. The key observation is that the CFT state on $S^2 \times \mathbb{R}$ can be obtained by inserting its corresponding operator into the vacuum at $\tau = -\infty$. Thus, for a scalar primary $\phi$, we have
\begin{equation}
    \langle 0| \phi(\mathbf{x}) | \phi \rangle = \frac{1}{R^{\Delta_\phi}}.
\end{equation}
This allows us to define a dimensionless two-point correlator as
\begin{equation}
    \frac{\langle 0| \phi(\mathbf{x}_1) \phi(\mathbf{x}_2)| 0\rangle }{\langle 0| \phi(\mathbf{x}) | \phi\rangle^2 } = \frac{1}{\left(2\sin(\theta_{12}/2)\right)^{2\Delta_\phi}}.
\end{equation}

In our fuzzy sphere model, we can similarly compute this dimensionless two-point correlator~\cite{Han2023Conformal}:
\begin{equation}
    G_{\sigma\sigma}(\theta_{12}) = \frac{\langle 0| n^z(\mathbf{x}_1) n^z(\mathbf{x}_2) |0\rangle }{\langle 0| n^z(\mathbf{x}) | \sigma\rangle^2}.
\end{equation}
Figure~\ref{fig:correlator} presents $G_{\sigma\sigma}(\theta_{12})$ for different values of $N = 2s+1$, showing excellent agreement with the theoretical prediction. One can use this dimensionless two-point correlator to accurately determine the critical point, e.g., by comparing $G_{\sigma\sigma}(\theta_{12} = \pi)$ with $2^{-2\Delta_\sigma}\approx 0.487577$, as shown in Figure~\ref{fig:correlator}c. Similarly, we can also compute the four-point correlator via $\langle \sigma| n^z(\mathbf{x}_1) n^z(\mathbf{x}_2) |\sigma\rangle$ as discussed in Reference~\citenum{Han2023Conformal}.

It is worth highlighting that the correlator in the fuzzy sphere model is a continuous function of $\theta_{12}$, demonstrating that the fuzzy sphere retains a continuous space structure, in contrast to the discrete nature of lattice models. More concretely, the correlator obtained on the fuzzy sphere is a truncated series expansion of $\cos^2(\theta_{12}/2)$ up to degree $2s$:
\begin{equation}
    G_{\sigma\sigma}(\theta_{12}) = \sum_{n=0}^{2s} a_n \cos^{2n}\frac{\theta_{12}}{2},
\end{equation}
which approximates the series expansion of the CFT two-point correlator,
\begin{equation}
    \frac{1}{\left(2\sin(\theta_{12}/2)\right)^{2\Delta_\sigma}} = 2^{-2\Delta_\sigma} \sum_{n=0}^{\infty} \frac{(\Delta_\sigma)_n}{n!} \cos^{2n}\frac{\theta_{12}}{2},
\end{equation}
where $(\Delta)_n$ is the Pochhammer symbol. This reveals a natural analogy between fuzzy sphere regularization and standard regularization schemes, in which a truncation at $\ell = 2s$ is introduced in the angular momentum. Indeed, this connection goes beyond mere analogy, and we refer to Reference~\citenum{He2025Gaussian} for further discussion.

It is remarkable that physical quantities, such as operators and their correlators, are well defined in real space in the fuzzy sphere model. Nevertheless, one might wonder whether this conflicts with the fact that we are studying models on a noncommutative geometry, in which coordinate uncertainty should arise due to the uncertainty principle. So what is the consequence of noncommutative geometry?  The noncommutative nature is indeed reflected in the fact that (equal time) quantum fields in the fuzzy sphere model generally do not commute at finite $s$, meaning that
\begin{equation}
    [n^\alpha(\mathbf{x}), n^\beta(\mathbf{y})] \neq 0 \quad \text{even for} \quad \mathbf{x} \neq \mathbf{y}.
\end{equation}
This differs from standard QFT, in which, for example, the canonical quantization condition imposes
\begin{equation}
    [\phi(\mathbf{x}), \pi(\mathbf{y})] = i \delta(\mathbf{x} - \mathbf{y}).
\end{equation}
One can interpret this noncommutativity as a form of nonlocality in the fuzzy sphere model. Importantly, this nonlocality (noncommutativity) vanishes in the limit $s \to \infty$, in which a commutative space is recovered.

An OPE coefficient can also be computed in a fashion similar to that of the dimensionless two-point correlator. On the CFT side, an OPE coefficient (of scalar operator) can be computed by
\begin{align}
	f_{ijk} = \frac{ \langle \phi_i| \phi_j(\mathbf x)|\phi_k\rangle }{ \langle 0|\phi_j(\mathbf x)|\phi_j\rangle }.
\end{align}
So one can choose different CFT states obtained in the fuzzy sphere model as well as the density field $n^z(\mathbf x)$ and $n^x(\mathbf x)$ to compute OPE coefficients involving $\sigma$ or $\epsilon$. Following this strategy, Reference~\citenum{hu2023operator} computes 13 OPE coefficients, some of which were previously unknown.

\subsection{Conformal Generators}\label{sec:generator}

In Section~\ref{sec:stateop}, we identify each conformal multiplet—i.e., a primary operator and its descendants—based on the integer spacings of their scaling dimensions. This integer spacing is a direct consequence of the conformal algebra, as discussed in Section~\ref{subsec:cft}. An important next step is to explicitly construct the conformal generators on the fuzzy sphere. This has been accomplished in References~\citenum{fardelli2024} and~\citenum{fan2024}, which we briefly review here.

In the fuzzy sphere model, we have manifest $SO(3)$ symmetry and dilatation (i.e., time translation). The missing generators are the translation generator $P_\mu$ and the special conformal transformation generator $K_\mu$, which emerge in the IR. The approach in References~\citenum{fardelli2024} and~\citenum{fan2024} is to use the time component of the stress tensor, $T_{\tau\tau}(\mathbf{x})$, which, up to a proper normalization, corresponds to the Hamiltonian density $H(\mathbf{x})$ of the fuzzy sphere model. Specifically, one has~\footnote{A similar idea was studied in 2D CFT~\cite{KOO1994,Milsted2017}, in which one can also construct the Virasoro generators in a similar way.},
\begin{equation}
    \frac{P_\mu + K_\mu}{2} = \int_{S^2} d^2 \mathbf{x} \, \mathbf{x}_\mu T_{\tau\tau} (\mathbf{x}).
\end{equation}
Furthermore, by utilizing the conformal algebra relations $[D, P_\mu] = P_\mu$ and $[D, K_\mu] = -K_\mu$, one can explicitly construct $P_\mu$ and $K_\mu$ on the fuzzy sphere. Using the 3D Ising fuzzy sphere model as an example, Reference~\citenum{fardelli2024} numerically verified many nontrivial predictions regarding the conformal generators, confirming the emergence of conformal symmetry in the fuzzy sphere Ising model.
Additionally, the generator $K_\mu$ can be used to distinguish primary operators, whereas $P_\mu$ helps identify the descendants of a given primary. This is particularly useful in cases in which two states have closely spaced energy levels.

\subsection{F-Function: Three-Dimensional Analog of the Central Charge in Two-Dimensional Conformal Field Theory}

RG theory is fundamental for understanding scale-dependent behaviors in critical phenomena. Because RG transformations integrate out short-distance degrees of freedom, a key feature of RG flow is its irreversibility—certain complexity measures, such as the number of degrees of freedom, monotonically decrease along the flow. The first established RG irreversibility theorem is Zamolodchikov's $c$-theorem in two dimensions~\cite{Zamolodchikov1986}. The $c$-theorem states that a $c$-function, associated with the central charge of 2D CFTs at RG fixed points, monotonically decreases under RG flow. The 2D $c$-theorem was later conjectured by Cardy~\cite{CARDY1988} to generalize to the $a$-theorem in four dimensions, which was subsequently proven by Komargodski \& Schwimmer~\cite{Komargodski2011}.  Both the 2D central charge and the 4D $a$-function are related to the conformal anomaly and can be computed from stress tensor correlators.

In three dimensions, a similar RG irreversibility theorem exists—the $F$-theorem. In this case, the RG monotonic $F$-function is defined through the partition function on the three-sphere $S^3$~\cite{Jafferis2011}. Mapping $S^3$ to $\mathbb{R}^3$ via a conformal transformation, the $F$-function appears in the entanglement entropy—specifically, it corresponds to the subleading term in addition to the conventional entanglement area law~\cite{Myers2010Seeing}. The entanglement version of the $F$-theorem has been proven using entanglement subadditivity~\cite{Casini2012Renormalization}. Unlike the $c$- and $a$-functions, the $F$-function is inherently nonlocal, as it is not related to the conformal anomaly and cannot be computed from the correlators of any local operators. Instead, it must be extracted from a nonlocal quantity on a conformally flat manifold, such as the entanglement entropy in $\mathbb{R}^3$ or $S^2\times \mathbb R$ (but not on $T^3$ or $T^2 \times \mathbb{R}$). This makes the $F$-function particularly challenging to compute.

The fuzzy sphere offers several unique advantages for computing the $F$-function~\cite{hu2024F}. On the fuzzy sphere, one can consider the entanglement entropy of a real-space entanglement cut at a latitude parameterized by $\theta$ on a sphere of radius $R$:
\begin{equation}\label{eq:arealaw}
    S_A (\theta) = -\textrm{Tr}(\rho_A \ln \rho_A) = \frac{\alpha R}{\delta} \sin\theta - F,
\end{equation}
where $\delta$ is a UV regulator. The first term represents the entanglement area law, whereas the second term corresponds to the $F$-function.  Because the space is continuous, we can define the cylinder-renormalized entanglement entropy~\cite{Banerjee2016} as
\begin{equation}\label{eq:cREE}
    \mathcal{F}_C (R, \theta_0) \equiv (\tan \theta \, \partial_{\theta} -1) S_A(\theta) \Big|_{R,\theta_0}.
\end{equation}
In the thermodynamic limit, where $R\sin\theta_0 \to \infty$, $\mathcal{F}_C(R,\theta_0)$ approaches $F$ at the IR fixed point.

Following this strategy, Reference~\citenum{hu2024F} reported the first nonperturbative computation of the $F$-function for the 3D Ising CFT, yielding $F_{\text{Ising}} = 0.0612(5)$. This value is slightly smaller than the $F$-function of a free real scalar, $F_{\text{free}} = \frac{\log 2}{8} - \frac{3\zeta(3)}{16\pi^2} \approx 0.0638$~\cite{Klebanov2011}, in agreement with the $F$-theorem. The result is also very close to predictions from the $4-\epsilon$ expansion, which gives $F_{\text{Ising}} \approx 0.0610$~\cite{Giombi2015} and $F_{\text{Ising}} \approx 0.0623$~\cite{Fei2015,Giombi2024}.

\section{A Path to the Zoology of Conformal Field Theory}

In the previous section, we used the 3D Ising CFT as an example to illustrate the concept of fuzzy sphere regularization and demonstrate its application in exploring various novel aspects of critical phenomena. In this section, we review the progress in extending the fuzzy sphere framework to other universality classes.

\subsection{Defect Conformal Field Theory and Boundary Conformal Field Theory}

Defects and boundaries are ubiquitous in the real world, and in the realm of critical phenomena, they can give rise to new physics, including distinct defect and boundary universality classes. A well-known example of a defect in critical phenomena is the Kondo effect, the study of which played a pivotal role in the development of RG theory, revolutionizing modern physics.

Introducing a defect or boundary into a bulk CFT can trigger an RG flow toward a nontrivial IR fixed point on the defect or boundary, leading to a defect CFT (dCFT)~\cite{Billo2013,Billo2016defect} or a boundary CFT (bCFT)~\cite{CARDY1984surface,cardy1989boundary,cardy1991bulkboundary}. The dCFT~\footnote{The general properties of defects and boundaries are similar, and a boundary can be viewed as a special type of defect. Thus, in general descriptions, we simply refer to defects, with the understanding that the discussion also applies to boundaries.} exhibits a rich array of new physical phenomena.  First, a given bulk CFT can host multiple defect universality classes, making it particularly interesting to explore the landscape of defect fixed points. For each defect fixed point, the original conformal symmetry of the bulk is broken down to a smaller conformal symmetry on the defect. New operators exist on the defect, which can still be classified into primaries and descendants under the defect conformal group.  Furthermore, there is a nontrivial interplay between the defect and the bulk, allowing for interesting correlation functions between bulk and defect operators. Finally, RG monotonicity theorems also apply to defect RG flows; for instance, the $g$-theorem of the line defect has been proven for bulk CFTs in various dimensions~\cite{Affleck1991Degeneracy,Cuomo2022Flow,Casini2023gTheorem}.

\begin{figure}[t]
	\includegraphics[width=0.99\linewidth]{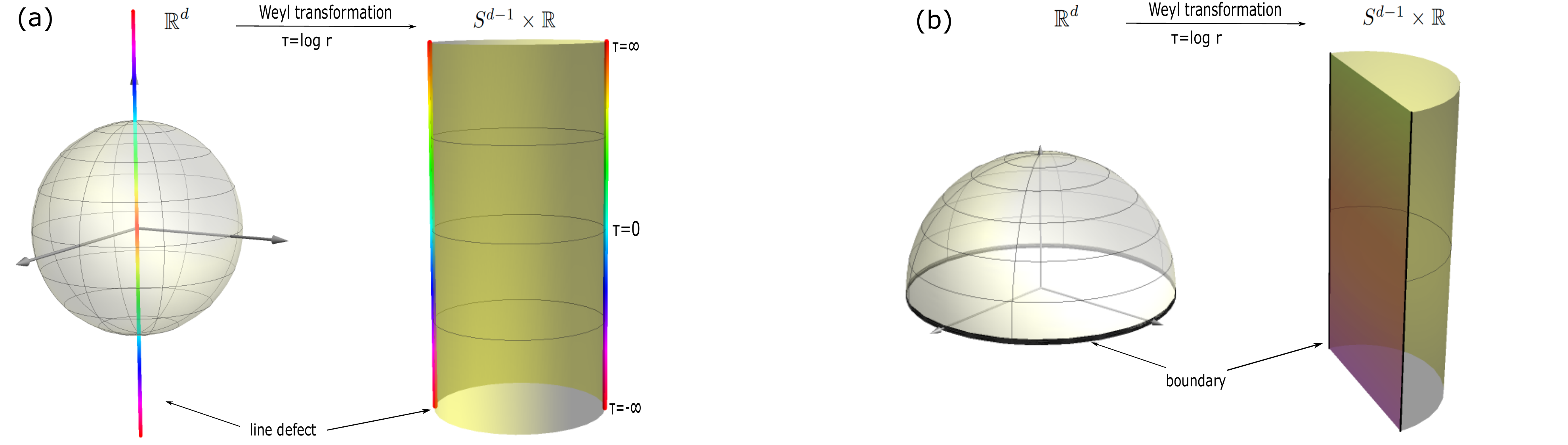}
		\caption{\textbf{Schematic plot of the defect and boundary in three dimensions}.
    (a) The line defect and (b) the boundary before and after the Weyl transformation.
	}
	\label{fig:drawing_defect}
\end{figure}

Similar to bulk CFTs, it is natural to study dCFTs in $\mathbb{R}^d$. Then by applying a Weyl transformation, one can map the theory onto the cylinder $S^{d-1} \times \mathbb{R}$, which again benefits from the advantages of the fuzzy sphere scheme. In particular, the state-operator correspondence remains valid for dCFTs.  As shown in Figure~\ref{fig:drawing_defect}, a line defect in $\mathbb{R}^d$ is mapped to $S^{d-1} \times \mathbb{R}$, where $0+1$D impurities are located at the north and south poles. Similarly, a CFT with a boundary in $\mathbb{R}^d$ is mapped to a setup in which the spatial dimensions reside on a hemisphere rather than a full sphere.  One can also consider defect-changing operators, which sit at the junction between two  different types of semi-infinite line defects. On $S^{d-1} \times \mathbb{R}$, this corresponds to placing different types of impurities at the north and south poles.

The simplest example of a conformal defect is the pinning field defect~\cite{Cuomo2022magneticDefect} in the 3D Ising CFT, in which an Ising $\mathbb{Z}_2$-breaking magnetic field is applied along a line defect. In field theory, this corresponds to turning on the $\sigma$ field on the 1D line defect, which is relevant ($\Delta_\sigma < 1$) and, therefore, induces an RG flow to a nontrivial IR defect fixed point.
In the fuzzy sphere model, this line defect can be realized by adding impurity terms at the north and south poles,
\begin{equation}
    h_{\textrm{N}} n^z(\mathbf{x} = \textrm{North Pole}) + h_{\textrm{S}} n^z(\mathbf{x} = \textrm{South Pole}),
\end{equation}
to the original Hamiltonian in Equation~\eqref{eq:ham_ising}. The eigenstates of this Hamiltonian are in one-to-one correspondence with the operators of the dCFT. Interestingly, different choices of impurity terms correspond to different types of operators: Setting $h_{\textrm{N}} = h_{\textrm{S}}$ yields defect operators, $h_{\textrm{N}} = -h_{\textrm{S}}$ gives defect-changing operators, and $(h_{\textrm{N}} \neq 0, h_{\textrm{S}} = 0)$ produces defect-creation operators. Reference~\citenum{hu2023defect} observed the state-operator correspondence for defect operators, explicitly demonstrated the defect conformal invariance, and obtained the scaling dimensions of several defect primary operators as well as bulk-defect OPE coefficients. Furthermore, Reference~\citenum{Zhou2024gfunction} computed the $g$-function using wave function overlap, obtaining $g = 0.602(2)$, which was a result later reproduced by the rigorous bootstrap approach (R. Lanzetta, S. Liu, and M. Metlitski, manuscript in preparation). In the same work, the authors also studied defect-creation and defect-changing operators, demonstrating that spontaneous Ising $\mathbb{Z}_2$ breaking is unstable on the line defect of the 3D Ising CFT. Additionally, Reference~\citenum{Cuomo2024} calculated the cusp anomalous dimension for the Ising pinning field line defect.

Shifting focus from defects to boundaries, References~\citenum{Zhou2025surface} and~\citenum{fuzzyHemi} studied the bCFT of the 3D Ising model on the fuzzy hemisphere. The key idea is that one can construct a fuzzy hemisphere by removing half of the Landau orbitals of the LLL. The authors investigated both the normal boundary (which explicitly breaks $\mathbb{Z}_2$ symmetry) and the ordinary boundary (which preserves $\mathbb{Z}_2$ symmetry). The emergent boundary conformal symmetry was once again demonstrated through the state-operator correspondence, along with results for boundary primary operators and bulk-boundary OPE coefficients. Furthermore, Reference~\citenum{Zhou2025surface} computed the boundary central charge, an RG-monotonic quantity governing the boundary RG flow~\cite{BoundaryC}. Additionally, the same work reported an intriguing connection between the bCFT operator spectrum and the bulk entanglement spectrum in Landau orbital space.

\subsection{$O(N)$ Wilson--Fisher Universality}

A natural generalization of the 3D Ising universality class is the $O(N)$ WF universality (e.g., see ~\cite{Vicari2002} for a review), where the Ising universality class corresponds to $N=1$. Similar to the Ising model, the $O(N)$ model describes an order--disorder transition, in which the ordered phase exhibits spontaneous symmetry breaking of $O(N)$. In nature, the $O(2)$ WF (also called XY) universality class describes the superfluid-to-normal fluid transition in helium-4, whereas the $O(3)$ WF universality class describes the Heisenberg ferromagnetic phase transition.

The general guiding principle for designing fuzzy sphere models is to ensure that the global symmetry and ('t Hooft) anomaly of the fuzzy sphere model match those of the target universality class. Because the $O(N)$ WF universality class does not exhibit any anomaly, the corresponding fuzzy sphere model must also preserve $O(N)$ symmetry and remain anomaly-free.
For the Ising universality class, we start with a model containing two fermion flavors, $\sigma = \uparrow, \downarrow$, whose maximal symmetry is $SU(2)$~\footnote{Strictly speaking, the global symmetry is $U(2)$ due to the additional $U(1)$ fermion number conservation.}. Next, we introduce interactions that break the $SU(2)$ flavor symmetry down to $\mathbb{Z}_2$, which is anomaly-free.
A sufficient but not necessary condition for being anomaly-free is the existence of a symmetric product state in the fuzzy sphere model. In the case of the fuzzy sphere Ising model Equation~\eqref{eq:ham_ising}, this state is the trivial paramagnet, $ \prod_{m=-s}^s (c^\dag_{m\uparrow} + c^\dag_{m\downarrow}) |0\rangle$.
This idea can be generalized to the $O(N)$ universality class. One starts with $N+1$ fermion flavors and introduces interactions that break the maximal $SU(N+1)$ symmetry down to $O(N)$. Here, the first $N$ flavors form an $O(N)$ vector, whereas the $(N+1)$-th flavor is an $O(N)$ singlet. When the fermion filling is one fermion per orbital, the fuzzy sphere model remains anomaly-free because one can construct a symmetric trivial product state, i.e., $\prod_{m=-s}^s c^\dag_{m,f=N+1} |0\rangle$.
Conversely, an $O(N)$ symmetry-breaking state corresponds to filling one of the first $N$ fermion flavors.
Indeed, the original Ising model is a special case of this generalized $O(N)$ model with $N=1$. Studies conducted in References~\citenum{Lauchli_XY} and~\citenum{Dey2025talk} and by W. Guo, Z. Zhou, T.C. Wei, and Y.-C. He (manuscript in preparation) have examined this type of $O(N)$ model up to $N = 4$. Additionally, Reference~\citenum{han2023o3} investigated a slightly different $O(3)$ WF model, which also involves four fermion flavors but with two fermions per Landau orbital~\footnote{The microscopic symmetry identification is also different between the two models.}.

\subsection{$SO(5)$ Deconfined Quantum Critical Point}\label{sec:dqcp}

The $SO(5)$ deconfined quantum critical point (DQCP)~\cite{Senthil2004a,Senthil2004b}, originally proposed to describe the phase transition between N\'eel magnetic order and valence bond solid~\cite{Sachdev1989}, is a key example of a phase transition beyond the conventional Landau symmetry-breaking paradigm. Since its proposal, the DQCP has been extensively studied (see ~\cite{Senthil2023DQCPReview} for a review), leading to numerous intriguing discoveries, including emergent symmetry~\cite{NahumSO5} and field theory dualities~\cite{WangChongduality}. However, its nature remains debated—lattice model simulations have observed anomalous scaling behavior~\cite{Nahum2015Deconfined}, leaving it unclear whether the DQCP is a weakly first-order transition or a continuous phase transition described by a CFT.

Reference~\citenum{zhou2023so5} studied the $SO(5)$ DQCP on the fuzzy sphere using one of its dual descriptions—a nonlinear sigma model on the target space $S^4$ with a Wess--Zumino--Witten (WZW) term. The same model has also been studied on the LLL of a torus~\cite{Ippoliti2018Half,Wang2021SO5WZW}, i.e., the fuzzy torus. The key discovery of Reference~\citenum{zhou2023so5} is that the energy spectrum of the $SO(5)$ DQCP exhibits an emergent conformal symmetry. Through the state-operator correspondence, numerous primary operators were also identified.  Furthermore, by analyzing the RG flow of the lowest $SO(5)$ singlet, the authors concluded that the DQCP may be pseudocritical (weakly first-order), with conformal symmetry likely emerging from nearby complex fixed points~\cite{WangChongduality,Gorbenko2018a,Gorbenko2018b,Kaplan2009}. However, direct evidence of these complex fixed points has yet to be found, leaving the debate on the nature of the DQCP unresolved. Follow-up studies on the fuzzy sphere~\cite{PhysRevB.110.125153,PhysRevLett.132.246503} reported similar numerical results but interpreted the DQCP as a tri-critical point, which is a proposal that was also made in References~\citenum{NSu2024} and~\citenum{takahashi2024} but still lacks a consistent theoretical framework.

\subsection{New Three-Dimensional $Sp(N)$ Conformal Field Theories}

The discovery of new 3D CFTs has long been a major pursuit, as it deepens our understanding of CFTs and facilitates the identification of new critical fixed points. The fuzzy sphere provides a promising platform for exploring such theories. As an example, Reference~\citenum{zhou2024newseries3dcfts} discovered a family of CFTs with $Sp(N)/\mathbb{Z}_2$ global symmetry, which generalizes the $SO(5) \cong Sp(2)/\mathbb{Z}_2$ DQCP.
The model considers $2N$ flavors of fermions with $2M$ flavors out of $2N$ to be filled. Interactions are then introduced to break the maximal $SU(2N)$ flavor symmetry down to $Sp(N)$. One can show that the effective low-energy description of this fuzzy sphere model is a nonlinear sigma model on the target space
\begin{equation}
    \frac{Sp(N)}{Sp(M) \times Sp(N-M)},
\end{equation}
supplemented by a level-1 WZW term. The $SO(5)$ DQCP corresponds to the case $(N,M)=(2,1)$.
Reference~\citenum{zhou2024newseries3dcfts} studied the cases $(N,M) = (2,1), (3,1), (4,1)$, and $(4,2)$ and in all instances, the authors observed emergent conformal symmetries, as evidenced by the integer-spaced conformal towers. Theoretically, the universality classes of these putative new CFTs remain unclear. Several candidate nonabelian gauge theories were proposed, which share the same global symmetry and anomaly with the fuzzy sphere model:
\begin{enumerate}
    \item $N$ flavors of critical bosons (in the fundamental representation) coupled to an $Sp(M)_1$ Chern--Simons gauge field.
    \item $N$ flavors of critical bosons (in the fundamental representation) coupled to an $Sp(N-M)_{-1}$ Chern--Simons gauge field.
    \item $N$ flavors of critical fermions (in the fundamental representation) coupled to an $Sp(1)_{N/2-M} \cong SU(2)_{N/2-M}$ Chern--Simons gauge field.
\end{enumerate}
Future study should focus on determining whether the fuzzy sphere model realizes one of these theories.

\subsection{Conformal Field Theory from Fractional Quantum Hall Transitions}

All the fuzzy sphere models discussed so far have focused on integer-filled Landau levels, meaning that the fermion number per Landau orbital is an integer. In this case, the electric charge of the fermions simply forms an integer quantum Hall state. Reference~\citenum{voinea2024} explored the case of fermions or bosons at fractional filling, in which the charge degrees of freedom instead form a more exotic fractional quantum Hall state. Interestingly, the authors observed high-quality conformal towers of the 3D Ising CFT at various fractional fillings, confirming the validity of the fractionally filled fuzzy sphere model.

Furthermore, the fuzzy sphere regularization scheme can be utilized to study the transitions between different quantum Hall states, opening a new avenue for exploring novel universality classes. In a fermion--boson mixture model, Zhou et al.~\cite{Zhou2025_2507} studied the confinement transition of the $\nu = 1/2$ bosonic fractional quantum Hall state, which was conjectured to have four different Lagrangian dual descriptions, such as one complex scalar boson coupled to a $U(1)_2$ Chern--Simons gauge field. The fuzzy sphere successfully demonstrates that this transition is continuous and conformal. It marks, for the first time, the fuzzy sphere as establishing a new conformal fixed point that was challenging to study by other methods. In addition, a similar fermion--boson mixture model can realize a free Majorana fermion field~\cite{Zhou2025_2509}. Voinea et al.~\cite{Voinea2025_2509} addressed a transition between the two-component abelian Halperin (221) state and a nonabelian Pfaffian state, wherein the universality is captured by a free Majorana fermion coupled to a $\mathbb{Z}_2$ gauge field~\cite{Wen2025}. To sum up, we expect that many types of universalities can be studied as quantum Hall transitions by the fuzzy sphere.

\subsection{Other Progress}
We conclude this section by briefly reviewing some more recent progress. Reference~\citenum{SYang2025} studied the three-state Potts model on the fuzzy sphere and observed numerical signatures of emergent conformal symmetry. The three-state Potts model has long been believed to exhibit a first-order transition~\cite{3DPotts_MC_1979,3DPotts_MCRG_1979,FERNANDEZ1989,JANKE1997,SWang2014}, suggesting that the conformal signature on the fuzzy sphere likely originates from nearby complex fixed points. The same phenomenon occurs in the 2D five-state Potts model, where this behavior is well understood both theoretically~\cite{Gorbenko2018a,Gorbenko2018b} and numerically~\cite{Tang2024,Wiese2024}. Future work should focus on searching for the complex fixed points of the three-state Potts model on the fuzzy sphere, which may also provide insights into resolving the nature of the $SO(5)$ DQCP, as the same physics has been conjectured to appear there.

References~\citenum{FanYangLee,Cruz2025,Miro2025} studied the 3D nonunitary Yang--Lee CFT on the fuzzy sphere, demonstrating that the fuzzy sphere scheme can effectively accommodate nonunitary CFTs. Additionally, References~\citenum{He2025Gaussian} and~\citenum{Taylor2025} established a concrete realization of the free scalar CFT on the fuzzy sphere, making progress toward formulating a general design principle for renormalizable Lagrangians in this framework.

\section{Outlook and Open Questions}

We have reviewed recent advances in fuzzy sphere regularization for studying critical phenomena. A key advantage of the fuzzy sphere model is its ability to extract a wealth of information with low computational cost. Looking ahead, we outline several intriguing directions for future exploration.

A key question in the fuzzy sphere scheme is how to construct a fuzzy sphere model that realizes a given universality class. So far, the guiding principle has been symmetry and anomaly matching. For instance, the $O(N)$ WF universality is studied using an $O(N)$-symmetric model in which $O(N)$ symmetry is anomaly-free. In contrast, the $SO(5)$ DQCP is modeled with an $SO(5)$-symmetric fuzzy sphere model whose anomaly is characterized by the Stiefel--Whitney class $w_4^{SO(5)}$ \cite{WangChongduality,LJZou2021}. Developing a systematic understanding of anomalies in fuzzy sphere models is crucial. With such a framework, we could realize a broader range of universality classes and address open problems in critical phenomena.

Another pressing question is how to better estimate numerical errors and optimize the accuracy of fuzzy sphere models. A brute-force approach is to improve numerical algorithms to access larger system sizes, thereby reducing finite-size effects. For instance, incorporating nonabelian symmetry quantum numbers (from both global symmetry and sphere rotation) into exact diagonalization and DMRG simulations would be highly beneficial. Also it is important to improve Monte Carlo simulation of fuzzy sphere models~\cite{Hofmann2024}. Another promising direction is to leverage CFT techniques, such as conformal perturbation theory \cite{Lao2023,Lauchli_CPT}, to refine numerical results and improve convergence.

From a broader conceptual viewpoint, fuzzy sphere regularization also resonates with Heisenberg’s original idea from the 1930s—employing noncommutative geometry to regularize QFTs to overcome their notorious UV divergences. Interestingly, Heisenberg’s proposal predated both the standard regularization methods of QFT and the formal mathematical establishment of noncommutative geometry by Connes. This early insight eventually evolved into noncommutative field theory \cite{noncommuQFT}, which examines QFTs defined in noncommutative spaces rather than conventional commutative ones. The early 2000s saw renewed interest in noncommutative field theory owing to its connection to D-branes in string theory. A significant finding was UV-IR mixing, which prevented Heisenberg’s idea from fulfilling its original promise, as noncommutative field theories turn out to be fundamentally distinct from standard field theories like $\phi^4$ theory or quantum electrodynamics. In contrast, the newly developed fuzzy sphere regularization successfully realizes standard QFTs using noncommutative geometry. More importantly, its unexpected numerical efficiency advocates a renewed exploration of Heisenberg's pioneering vision. To advance in this direction, it is helpful to explore the connections and differences between this framework and previous studies of noncommutative field theory on the fuzzy sphere~\cite{Madore2001Scaling}. A related question is whether there exists a systematic procedure for constructing renormalizable Lagrangians on the fuzzy sphere. Establishing such a framework would extend the applicability of the fuzzy sphere beyond fixed points and universality classes, enabling the study of RG flows and field theory dualities. Some initial discussions on these topics have been presented in Reference~\citenum{He2025Gaussian}.

We conclude this review by highlighting the following interesting directions for the fuzzy sphere scheme:
\begin{enumerate}
\item Developing theoretically tractable fuzzy sphere models for 3D CFTs would be highly valuable. This includes constructing exactly solvable models for free CFTs and large-$N$ models that can be studied perturbatively. For free theories, leveraging higher-spin currents may facilitate the construction of exactly solvable models. Meanwhile, fuzzy sphere models of the $O(N)$ WF and newly discovered $Sp(N)$ universality classes could serve as promising candidates for large-$N$ expansion computations.

\item The fuzzy sphere framework could also be used to discover new CFTs, such as the Stiefel liquid \cite{LJZou2021}, which has been proposed as a non-Lagrangian theory. The Stiefel liquid is closely related to the $SO(5)$ DQCP and $N_f = 4$ QED$_3$.

\item Another promising direction is using the fuzzy sphere to advance the study of conformal defects and boundaries. For example, an exciting avenue is to investigate Wilson and vortex loops in gauge theories.

\item Studying complex fixed points within the fuzzy sphere model is crucial. Potential targets include the Potts model and the $SO(5)$ DQCP. A deeper understanding of complex fixed points could help resolve debates surrounding the $SO(5)$ DQCP and shed light on the conformal window problem in gauge theories.

\item Exploring CFTs at finite temperature \cite{El_Showk_2012,Iliesiu_2018}, i.e., on $S^2 \times S^1$ rather than $S^2 \times \mathbb{R}$, is another important direction. This would provide insights into quantum criticality at finite temperatures, a key aspect for experimental studies.

\item Understanding the entanglement structure of fuzzy sphere models could offer new insights into the entanglement properties of 3D CFTs and explain the remarkable efficiency of the fuzzy sphere approach. A particularly intriguing question is the observed correspondence between the entanglement spectrum and the operator spectrum of bCFTs \cite{Zhou2025surface}.

\item Investigating the role of the Girvin--MacDonald--Platzman (GMP) algebra~\cite{GMP} in the fuzzy sphere scheme is another compelling direction. Reference~\citenum{He2025Gaussian} has shown that in certain limits, the (generalized) GMP algebra reduces to the harmonic oscillator algebra on the sphere. It would be interesting to explore whether the (generalized) GMP algebra is related to conformal algebra or reveals new algebraic structures in 3D CFTs.

\item The fuzzy sphere also provides an excellent platform for studying nonequilibrium dynamics and quantum chaos in 3D CFTs. Its high numerical efficiency makes it feasible to investigate these phenomena using exact diagonalization and DMRG techniques.

\item Finally, it is worth exploring generalizations of the fuzzy sphere scheme applied to other manifolds. A natural extension is studying CFTs on $S^4 \times \mathbb{R}$, where Landau levels of the $SU(2)$ monopole can be considered \cite{Zhang_2001}. A recent work also generalizes the fuzzy approach to $\mathbb{RP}^2$~\cite{Dong2025}.

\end{enumerate}

\clearpage

\section*{DISCLOSURE STATEMENT}
The authors are not aware of any affiliations, memberships, funding, or financial holdings that
might be perceived as affecting the objectivity of this review.

\section*{ACKNOWLEDGMENTS}
We thank Gabriel Cuomo, Davide Gaiotto, Chao Han, Johannes S. Hofmann, Emilie Huffman, Liangdong Hu, Zohar Komargodski, Chong Wang, Zheng Zhou, and Yijian Zou for collaborating on various projects.
This work was supported by the National Science Foundation of China under grant number 12474144 (W.Z.).

\bibliographystyle{ar-style4.bst}
\bibliography{ref_FSR.bib}

\end{document}